\documentclass[lettersize,journal]{IEEEtran}

\usepackage{amsmath,amsfonts}
\usepackage{array}
\usepackage[caption=false,font=normalsize,labelfont=sf,textfont=sf]{subfig}
\usepackage{textcomp}
\usepackage{stfloats}
\usepackage{url}
\usepackage{verbatim}
\usepackage{graphicx}
\def\BibTeX{{\rm B\kern-.05em{\sc i\kern-.025em b}\kern-.08em
    T\kern-.1667em\lower.7ex\hbox{E}\kern-.125emX}}
\usepackage{balance}
\usepackage{amssymb}
\usepackage{graphicx}      
\usepackage{epsfig} 
\usepackage{mathptmx} 
\usepackage{times} 
\usepackage{amsmath} 
\usepackage{booktabs} 
\usepackage{algorithm}
\usepackage{algpseudocode}
\usepackage{xcolor}
\usepackage{svg}

\begin{document}

\title{Neural Networks for on-chip Model Predictive Control: a Method to Build Optimized Training Datasets and its application to Type-1 Diabetes}

\author{A. Castillo, E. Pryor, A. El Fathi, B. Kovatchev, and M. Breton%
	\thanks{A. Castillo, E. Pryor, A. El Fathi, B. Kovatchev, and M. Breton are with the Center for Diabetes Technology, University of Virginia, US. Emails:}%
	\thanks{\texttt{cfv4mj, hyy8sc, fwt9vd, bpk2u, mb6nt}@virginia.edu}%
	\thanks{© 2025 IEEE. Personal use of this material is permitted. Permission from IEEE must be obtained for all other uses, in any current or future media, including reprinting/republishing this material for advertising or promotional purposes, creating new collective works, for resale or redistribution to servers or lists, or reuse of any copyrighted component of this work in other works.}
}

\markboth{04/13/2025 --- Preprint --- This article has been accepted for publication at the \textit{IEEE Transactions on Cybernetics}}%
{}

\maketitle

\begin{abstract}

Training Neural Networks (NNs) to behave as Model Predictive Control (MPC) algorithms is an effective way to implement them in constrained embedded devices. By collecting large amounts of input-output data, where inputs represent system states and outputs are MPC-generated control actions, NNs can be trained to replicate MPC behavior at a fraction of the computational cost. However, although the composition of the training data critically influences the final NN accuracy, methods for systematically optimizing it remain underexplored. In this paper, we introduce the concept of Optimally-Sampled Datasets (OSDs) as ideal training sets and present an efficient algorithm for generating them. An OSD is a parametrized subset of all the available data that (i) preserves existing MPC information up to a certain numerical resolution, (ii) avoids duplicate or near-duplicate states, and (iii) becomes saturated or complete. We demonstrate the effectiveness of OSDs by training NNs to replicate the University of Virginia’s MPC algorithm for automated insulin delivery in Type-1 Diabetes, achieving a four-fold improvement in final accuracy. Notably, two OSD-trained NNs received regulatory clearance for clinical testing as the first NN-based control algorithm for direct human insulin dosing. This methodology opens new pathways for implementing advanced optimizations on resource-constrained embedded platforms, potentially revolutionizing how complex algorithms are deployed.
\end{abstract}

\begin{IEEEkeywords}
Model Predictive Control, Machine Learning, Automated Insulin Delivery, Artificial Pancreas, Data Curation.
\end{IEEEkeywords}

\section{Introduction}
\IEEEPARstart{M}{odel} Predictive Control (MPC) is a type of feedback regulator that computes the control action by solving a constrained optimization problem \cite{mayne2014model}. Since its inception, MPCs have gained widespread adoption across various industries, primarily owing to their capacity to manage complex systems with multiple inputs and outputs \cite{yan2024data,wang2021model,huang2024data,liu2022model,yang2023disturbance}.

Despite their success, it is acknowledged that MPCs entail significant computational complexity. MPC requires solving a constrained optimization problem in each control cycle to determine the optimal control action. This complexity poses challenges to the feasibility of MPCs in applications where the control algorithm needs to run in small devices; such as smartwatches, pods, insulin pumps, or rings \cite{liegmann2021real}. For example, in Automated Insulin Delivery (AID) for type-1 diabetes, MPCs are showing promising results \cite{garcia2021advanced,Forlenza2022,ozaslan2022zone,sun2022personalized,hajizadeh2019plasma, ganji2024lmi}, but it remains unclear how they could be embedded inside commercially available insulin pumps without compromising other hardware requirements \cite{castle2017future}.

Recently, a new way to overcome the MPC computational complexity was studied in \cite{castillo2023deep,kivs2021explicit,maddalena2020neural,winqvist2020training,chen2018approximating}. The idea consists in training Neural Networks (NNs) with Rectified Linear Unit (ReLU) activations to replicate the MPC behavior. The NNs are trained with synthetically generated data pairs, $\{x_i,\,u_i\}$; where~$x_i$ represents all the parameters needed to solve the MPC problem and $u_i=MPC(x_i)$ is the resulting MPC control action. After training, the NN effectively replicates the MPC behavior while being remarkably efficient to compute. 
\begin{figure*}[t]
    \centering
    \includegraphics[width=6.15in]{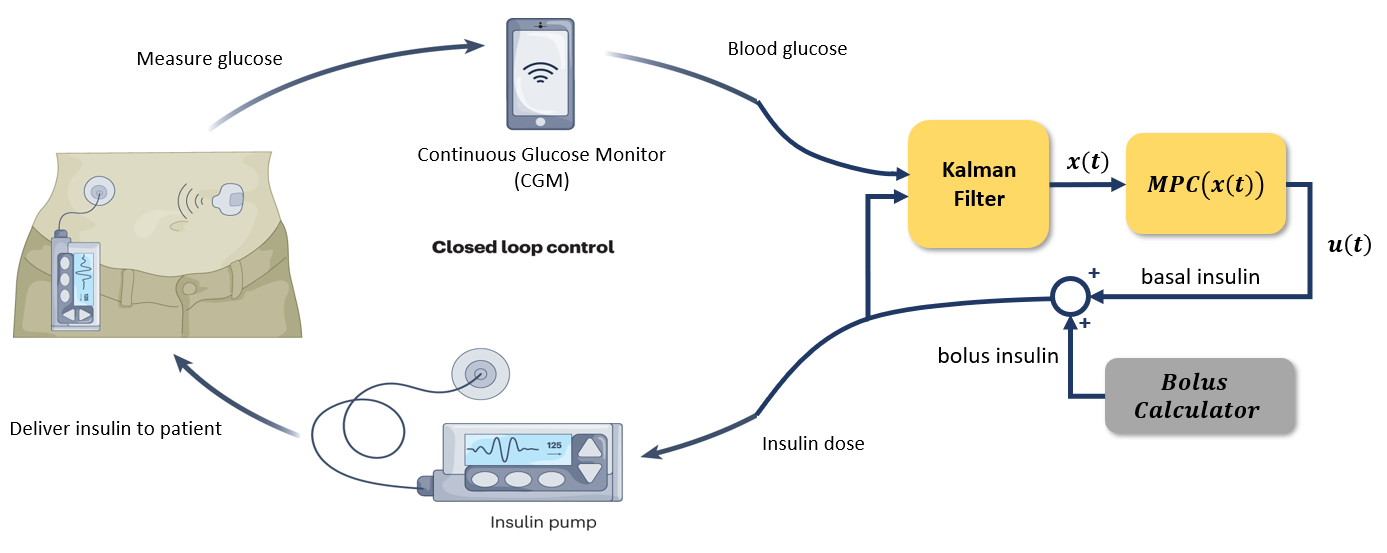}
    \caption{Simplified block diagram of the University of Virginia Automated Insulin Delivery System~\cite{garcia2021advanced}.}
    \label{fig:AID}
\end{figure*}

When analyzing deployments in wearable devices, key performance considerations include the memory footprint (comprising both non-volatile program memory and RAM data memory) \textcolor{black}{and the computational complexity measured by the number of operations required to obtain a solution.} In~\cite{kivs2021explicit}, it was demonstrated that encoding the closed-form MPC solution in NN weights achieves a significant 1 million-fold reduction in program memory, reaching orders of \textit{bytes} for simple MPCs. The work \cite{castillo2023deep} showed that Residual~NN architectures can further improve both computational efficiency and memory footprint, reducing execution to only a few low-order matrix multiplications. These results align with studies \cite{fahandezh2020proximity,montufar2014number,arora2016understanding} analyzing the relationships between ReLU-NNs, Piece-Wise Affine (PWA) functions, and the MPC problem. 

In consequence, evidence suggests that NNs can efficiently learn and replicate MPC behavior. As NNs are lightweight and require only a single deterministic forward pass to generate a solution, they can be deployed on small chips or embedded platforms where classical optimization algorithms may struggle due to computational, power, or memory constraints. This capability opens the door to obtain efficient on-chip MPC, a longstanding goal that has captured researchers' attention for decades \cite{schwenzer2021review,frison2014efficient, richter2011towards, kouzoupis2015towards, arnstrom2023bnb, frison2014high, cimini2020embedded}.

However, despite these results, the development of methods to create optimized training datasets for MPC-NNs remains a significant unaddressed challenge \cite{winqvist2020training}. Training neural networks with unbalanced, incomplete, or unstructured data can introduce critical inaccuracies and compromise predictable performance. Consequently, developing algorithms to optimize and manage the training sets is paramount, as these methodologies ultimately enhance the reliability of this strategy.

In this paper, we define the notion of Optimally Sampled training Dataset (OSD), which are training sets characterized by three properties: \textbf{i)} they contain no repeated elements; \textbf{ii)} they achieve adaptive granularity and numerical resolution via a cost function; and \textbf{iii)} they can reach a complete or saturated state. We develop a scalable algorithm with $\mathcal{O}\bigl(\log(N_d)\bigr)$ complexity (where $N_d$ is the dataset size) to create OSDs from vast amounts of unstructured data, and apply it to the problem of insulin regulation in Type-1 Diabetes (T1D) management. We show that NNs trained on OSDs achieve consistently higher accuracy compared to those trained on conventional approaches. Lastly, we discuss how this methodology could be extended to other application domains.

\section{Motivations}
This work was driven by two motivations:

\subsection{Automated Insulin Delivery for Type-1 Diabetes.}\label{sec:UVAMPC}
Advancements in AID systems for T1D, such as the research developed at the University of Virginia~\cite{patek2009silico, patek2012modular, breton2012fully, kovatchev2016artificial, brown2019six, garcia2019closed, garcia2021advanced}, have demonstrated significant potential in enhancing patient care. These systems increasingly employ Model Predictive Control (MPC) algorithms to determine insulin infusion rates every 5 minutes based on Continuous Glucose Monitoring (CGM) data, as illustrated in Fig.\ref{fig:AID}. MPCs offer promising features by incorporating adaptive glucose-insulin metabolic models, efficiently encoding system delays, and managing input constraints. The ultimate objective is to transfer insulin administration and glucose regulation responsibilities from the individual to the algorithm, thereby improving both glycemic control and patient’s quality of life.

Medical devices should be as unobtrusive to patients as possible, leading to deploying the AID control algorithm inside wearables, such as insulin pumps, pods or smart watches; which are low-energy and increasingly small devices that share hardware and energy sources with other mechanical/electrical elements. Implementing intensive algorithms in wearable medical devices could easily impact their core design constraints, such as responsiveness, battery life, or temperature~\cite{castle2017future}.

This work addresses the need for efficient implementation of MPC algorithms in highly constrained environments, with the goal of leveraging the advantages of MPC control in T1D without imposing limitations in the target medical devices.




\subsection{Certified Implementations of Model Predictive Controllers}
Since the early development of the MPC technology, questions regarding the accuracy and safety of its implementation have persisted. Most MPC algorithms lack a manageable closed-form formula and, therefore, they cannot be deployed in an exact manner through mathematical equations. 

To overcome this problem, researchers developed search algorithms (also known as MPC solvers) that approximately find a solution in an iterative way. Industries and regulatory agencies have largely accepted MPC solvers as the gold standard for practical implementations due to their `certifications'; meaning that mathematical methods were employed in order to prove useful solver properties, such that the difference between $u^*$ (the ideal MPC control action) and~$u_s$ (the one found by the solver) will be smaller than a certain tolerance.

This type of certifications enabled industries and regulatory agencies to trust the technology, contributing to make the MPC one of the most popular feedback control methods. Hence, similar results are necessary when the implementation artifact is an NN instead of an iterative search algorithm. The OSDs presented in this paper may offer a formal certification of the quality of the data used to train the network.

\section{Model Predictive Control}\label{sec:MPC}
Model Predictive Control (MPC) is a strategy for designing feedback in a dynamical system. It uses an internal state-space model, $x_{k+1} = f(x_k, u_k)$, together with an available initial state observation, $x_0=x(t)\in\mathbb{R}^n$, to generate an optimal sequence of control actions, $u_0,\dots,u_{N_c-1}$, $u_k\in\mathbb{R}^m$, that minimizes a given cost $J$. An MPC is therefore defined as:
\begin{equation} \label{eq:MPC}
\begin{aligned}
\min_{u_0,\dots,u_{N_c-1}} \quad & J=\sum_{k=0}^{N_c-1} L(x_k, u_k) + M(x_{N_c}) \\
\text{subject to} \quad & x_{k+1} = f(x_k, u_k); \; x_0 = x(t); \\
& g(x_k, u_k) \leq 0;\, h(x_k, u_k) = 0;\\
\end{aligned}
\end{equation}
where $L(\cdot)$, $M(\cdot)$ are cost functions, $g(x_k, u_k) \leq 0$ and $h(x_k, u_k) = 0$ are constraints, and $N_c$ is the prediction horizon.

After solving \eqref{eq:MPC} and getting the optimal control sequence, $u_0,\dots,u_{N_c-1}$, only the first input, $u(t)=u_0$, is applied to the system and then the process is re-run at the next time-step. 

Due to this property, an MPC can be simply understood as a deterministic input-output function:
\begin{equation} \label{eq:MPCnn}
u(t) = MPC(x(t)),
\end{equation}
where $x(t)$ represents the system-state at time-step `$t$', while $u(t)$ represents the first element of the optimal control sequence found by the solver.

\begin{figure*}[t]
    \centering
    \includegraphics[width=6.15in]{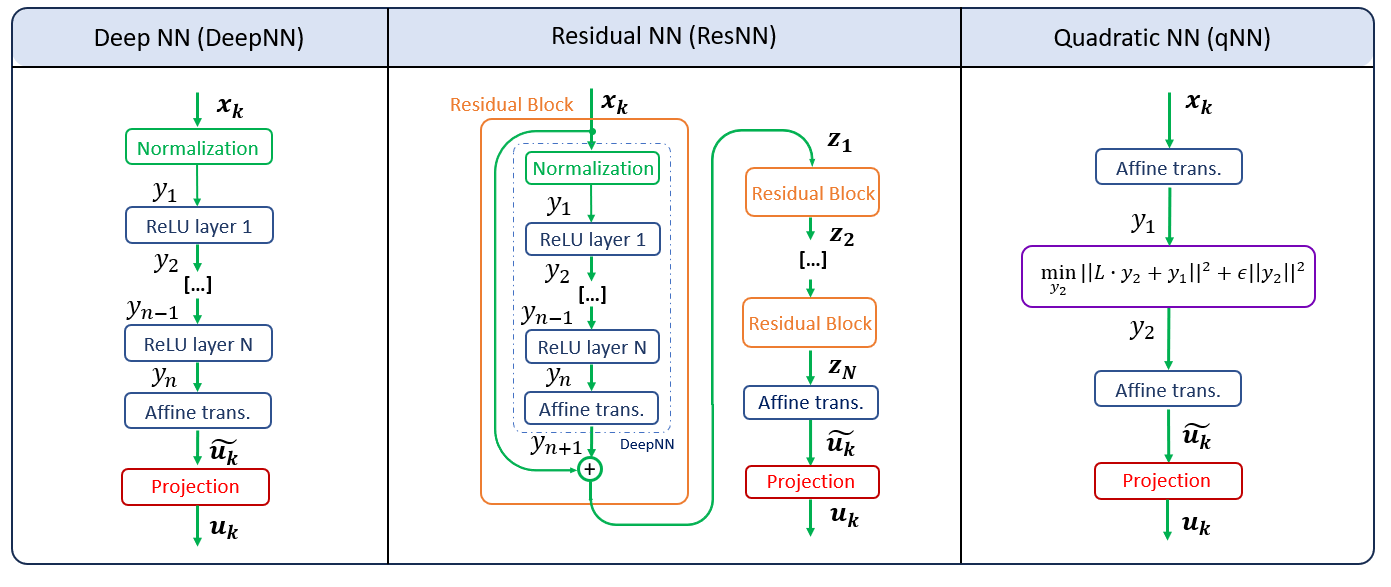}
    \caption{Three main NN architectures that have been proposed for MPC learning. The individual operators are: Normalization$:=\frac{x_k-\mu}{\sqrt{\sigma}+\epsilon}$, ReLU layer$:=\max\{0,\,(A_{k}\cdot y_{k}+b_{k})\}$, Affine trans.$:=A_{k}\cdot y_{k}+b_{k}$, and Projection$:=\min_{u_k}||u_k-\tilde{u_k}||^2$, s.t. $C_uu_k\leq d_u$ and $C_cBu_k\leq d_c-C_cAx_k$; where $C_u$, $C_c$, $d_u$ and $d_c$ are defined in \cite{chen2018approximating} and are designed to guarantee the MPC constraints.}
    \label{fig:ResNET}
\end{figure*}

\subsection{Embedded implementation issues}
Implementation issues arise because a manageable closed-form solution for \eqref{eq:MPC} rarely exists. 

If the model $x_{k+1} = f(x_k, u_k)$ is linear and the costs $L(x, u)$, $M(x)$ are quadratic, a closed-form solution was proved to exist. In that case, the Eq. \eqref{eq:MPCnn} becomes a continuous Piece-Wise Affine (PWA) function defined over polytopic sets \cite{alessio2009}. Even though the PWA function can be calculated in advance, the number of sets often becomes excessively large, complicating embedded implementations due to memory constraints.

Thus, MPCs are generally implemented with solvers \cite{schwenzer2021review}. There are two main building blocks for MPC solvers: first-order \cite{richter2011towards} and second-order (or interior-point) methods \cite{domahidi2012efficient}. First-order methods are based on executing gradient descent steps followed by projections onto the constrain feasible set, while second-order methods make use of the second-order information in the search direction computation \cite{frison2014efficient}.

Deploying MPC solvers in embedded devices poses three main problems, which have led to different improvements over the years \cite{frison2014efficient, richter2011towards, kouzoupis2015towards, arnstrom2023bnb, frison2014high}:
\begin{enumerate}
    \item \textbf{Intense algebra.} Computing first/second order gradients plus projections several times per period is expensive and may pose concerns in low-energy devices. 
    \item \textbf{Uncertain number of cycles.} The number of iterations to find a solution is sensitive to the initial condition and remains uncertain, introducing variability in battery life. Although upper bounds can be established (falling into the `performance certification' analyses mentioned above \cite{richter2011towards, arnstrom2023bnb}), they tend to be conservative.
    \item \textbf{Solver compilation for the target device.} The solver needs to be compiled and transferred to the target device, which will need to be compatible and have enough memory to store the program.
\end{enumerate}

\subsection{Neural Networks to deploy MPCs}
Recently, NNs have been suggested for on-chip MPC deployments. 
The idea is to build a new pipeline, consisting first on using solvers to generate synthetic training data, and then using the data to train NNs for final deployment. The training data consists of a set of pairs $\{x,u\}$, from which the NN adjusts its weights to learn the function \eqref{eq:MPCnn} that generated it.

Figure~\ref{fig:ResNET} contains the three main architectures that have been proposed for MPC learning. The first architecture (left), was considered in \cite{chen2018approximating} and it represents a standard deep NN with Rectified Linear Unit Activations (ReLU). The second architecture (middle) was considered in \cite{castillo2023deep} to minimize computational loads, providing significant reductions in parameters without sacrificing accuracy. The third architecture (right) was introduced in~\cite{maddalena2020neural} and it fits a simplified quadratic cost that approximates~\eqref{eq:MPC} through the available dataset. 

Any of the three architectures is trained on a synthetically generated data set:
\begin{equation}\label{eq:training_set}
    \mathcal{D} = \big\{X_0,\, X_1,\, ...\,, X_{N_d}\big\},\quad X_t\triangleq\{x,u\}\in\mathbb{R}^{m+n},
\end{equation}
where $X_t$ represents a state observation together with its optimal control action \eqref{eq:MPCnn}, and $N_d$ is the dataset size.

During the training process, the NN adjusts its internal weights to minimize a cost, normally of the form:
\begin{equation}\label{eq:MSE_optimized_no_bracket}
    \theta^* = \underset{\theta}{\text{arg min}} \left\{\frac{1}{N_d+1} \sum_{t=0}^{N_d} \left[\tilde{u}(x,\theta) - u\right]^2 \right\}
\end{equation}
where $\tilde{u}(x,\theta)$ represents the NN function, $\theta$ all its learnable parameters, and $u$ the target control action \eqref{eq:MPCnn}.

\begin{figure*}[t]
    \centering
    \includegraphics[width=6.5in]{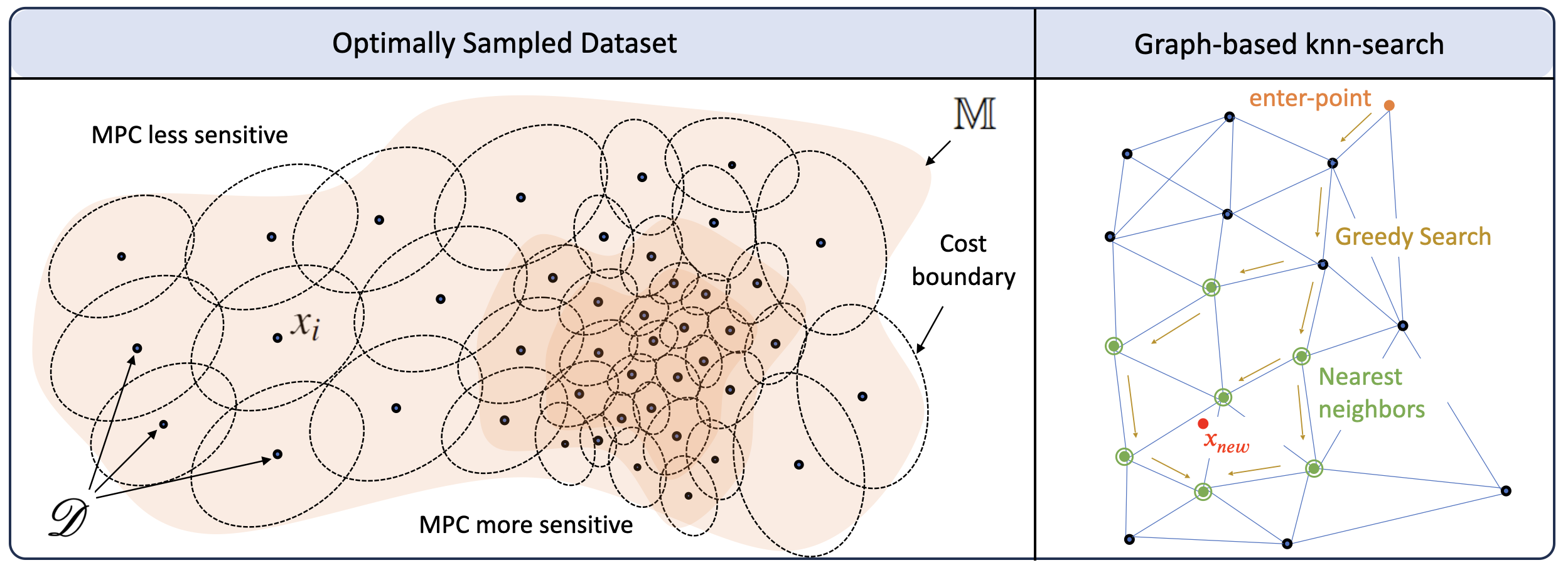}
    \caption{\textbf{Left} --- Illustration of an Optimally Sampled Dataset (OSD). The MPC operating space, $\mathbb{M}$, is partitioned in a finite set of discrete volumes. The volumes automatically shrink thanks to the control action penalization term in order to augment sampling and resolution in regions where the MPC becomes more sensitive; i.e. higher $\partial(MPC(x))/\partial x$. The space inside each volume remains empty, while the final dataset $\mathcal{D}$ is conformed by the centroids. \textbf{Right} –-- Fast search algorithm for nearest neighbor retrieval~\cite{malkov2018efficient}. Every centroid is connected to its neighbors. The nearest neighbor search is initiated at a random entry-point, iteratively moving to the next neighbors that makes the cost smaller.}
    \label{fig:D}
\end{figure*}

\subsection{The importance of having balanced training sets}\label{sec:training_sets}
The quality and distribution of the training data play a crucial role in the final NN performance. For this reason, it is essential to ensure that the training data distribution comprehensively captures all the control mapping $MPC(x)$ with sufficient resolution and granularity.

On the one hand, training on raw unfiltered data may lead to errors and local overfitting due to a skewed probability data distribution. Statistically more common states will appear more frequently, making the cost-minimization \eqref{eq:MSE_optimized_no_bracket} unbalanced. Consequently, the neural network may over-perform in prevalent states and under-perform in rare states. On the other hand, a uniform sampling throughout the entire input state-space (as proposed in previous works \cite{winqvist2020training}) might not be ideal because MPC solutions are non-uniform. Such non-uniformity in the MPC solution necessitates increased sampling and finer resolution in critical areas to prevent information loss.

For these reasons, the problem of how to generate suitable training sets for MPC learning becomes paramount. In the next sections, the concept of \textit{Optimally Sampled Dataset (OSD)} is defined, providing a new useful mechanism for building datasets that minimize both issues.

\section{Optimally Sampled Training Datasets}\label{sec:SODD}
The MPC operates on states, $x$, that belong to a continuous subspace of $\mathbb{R}^n$; i.e. ${x\in\mathbb{M}\subseteq \mathbb{R}^n}$. The subspace $\mathbb{M}$ is referred to as the MPC `operational space', whose shape and size depend on the model matrices and the disturbances that are handled during operation. This space can be equal to $\mathbb{R}^{n}$ but, in practice, it is significantly smaller as not all states can be reached due to limitations in the actuators and disturbances.

The goal is to define an optimized discrete sampling, $\mathcal{D}$, of the continuous MPC operational space, $\mathbb{M}$; which will be the final NN training set \eqref{eq:training_set}. This is defined in three steps:

\vspace{0.1cm}
\textbf{Definition 1 --- Quadratic cost:} For a pair of states, $x_i,\,x_j\in\mathbb{M}$, a cost is computed as:
\begin{equation}\label{eq:cost}
J(X_i,X_j) = (x_i-x_j)^T S_x (x_i-x_j) + (u_i-u_j)^T S_u (u_i-u_j),
\end{equation}
being $S_x,\,S_u$ positive definite matrices and $u=MPC(x)$.

\textbf{Definition 2 --- }$\mathcal{D}$\textbf{-Nearest Neighbor:} Given an state, $x\in\mathbb{M}$, its $\mathcal{D}$-Nearest Neighbor is another state, $x^{nn}$, existing in the dataset \eqref{eq:training_set} that minimizes the cost \eqref{eq:cost}:
\begin{equation}
x^{nn} = \underset{\forall\;x_i \in \mathcal{D}} {\arg \min} \; J(x_i,x)
\end{equation}

\textbf{Definition 3 --- Optimally Sampled Dataset (OSD):} A dataset \eqref{eq:training_set} is an optimal sample of $\mathbb{M}$ with density~$J^*$ and accuracy $u_s$ if: \textbf{i)} For every $x\in \mathbb{M}$, we can find a $\mathcal{D}$-Nearest Neighbor, $x^{nn}$, with cost lower than $J^*$ satisfying $|MPC(x)-MPC(x^{nn})|\leq u_s$; \textbf{ii)} For every state $x_{i}$ in the dataset \eqref{eq:training_set}, we cannot find a $\mathcal{D}$-Nearest Neighbor with cost lower than~$J^*$.

\vspace{0.25cm}
The first definition establishes a cost that, starting from zero when $x_i=x_j$, it grows as both states are separated from each other. The matrices $S_x$ and $S_u$ define how the cost grows, where $S_x$ defines a normalized distance metric for the system states, while $S_u$ penalizes differences in their control actions.

The second definition establishes the dataset nearest neighbor notion (or $\mathcal{D}$-Nearest Neighbor). If a state $x\in\mathbb{M}$ is given, then its $\mathcal{D}$-Nearest Neighbor is the state in \eqref{eq:training_set} minimizing $J$.

With the $\mathcal{D}$-Nearest Neighbor notion, the optimized discrete sampling is established in Def.~3. In simple words, if we randomly choose a state $x\in\mathbb{M}$ from the MPC operational space, then we will always find a $\mathcal{D}$-Nearest Neighbor in our training set \eqref{eq:training_set} with a cost lower than $J^*$ and control action error lower than $u_s$ (first point of Def.~3). This means that the training set \eqref{eq:training_set} is complete up to a certain numerical resolution defined by $J^*$. The second point states that, if we randomly choose a state $x_i$ from the training dataset \eqref{eq:training_set}, then we will not find any other $\mathcal{D}$-Nearest Neighbor with cost lower than $J^*$. This implies that each state in the training set is unique, with no state repetitions within distances lower than $J^*$. 

This cost-based definition basically creates a partition of the continuous MPC operational space $\mathbb{M}$ in a finite set of discrete volumes, as illustrated in Figure~\ref{fig:D}. Each element $x_i \in \mathcal{D}$ creates an empty volume defined by: 
\begin{equation}\label{eq:partition_volume}
    \mathcal{E}_i=\{x\in\mathbb{M}\;\vert \; J(x_{i}, x) \leq J^*\},
\end{equation}
and all volumes $\mathcal{E}_i$ fill the complete space $\mathbb{M}$.

The interesting property is that the cost provides a simple mechanism to automatically adjust the size and shape of these partition-volumes $\mathcal{E}_i$. The boundary $(x_k-x)^TS_x(x_k-x)=J^*$ defines an initial maximum-size ellipsoid, which is then shrunk in a non-uniform way due to the control action penalization term: $(x_k-x)^TS_x(x_k-x)+(u_i-u)^T S_u (u_i-u)=J^*$. This property permits to automatically adjust the size and shape of the partition volumes in order to minimize information loss, augmenting the granularity and density in regions where the MPC function, $u=MPC(x)$, becomes more sensitive to small changes in the system state, $x$. Some illustrations are given in Figure~\ref{fig:D} and in Section~\ref{sec:results}.

\section{An algorithm to build OSDs from unstructured data: application to Diabetes Control}\label{sec:OSD_alg}
In this section, we propose a method to construct OSDs from extensive unfiltered data. Starting from an initial dataset of state-action pairs, $\mathcal{D}_o$, which effectively captures the MPC operational space (i.e., $\mathcal{D}_o \approx \mathbb{M}$), the objective is to refine this dataset to a minimal final one that meets the previously defined OSD criteria. An algorithm is introduced to this purpose.

The approach is tailored to the specific use-case of automated insulin delivery for T1D, where an MPC determines optimal insulin dosages based primarily on blood glucose sensor inputs. The UVA-MPC algorithm is used as a reference, but the methodology can be applicable to other scenarios.

\subsection{Target MPC description}\label{sec:UVAMPC_description}
The UVA AID system is highlighted in Fig.~\ref{fig:AID} and its MPC responds to the general formulation \eqref{eq:MPC}. First, a state-space model $x_{k+1}=f(x_k,u_k)$ is defined in order to represent the glucose-insulin dynamics. The current system-state estimation $x_0 = x(t)$ is obtained with a Kalman filter based on the recent history of insulin delivery and blood glucose. This state is used by the MPC to compute future glucose predictions and optimize insulin dosing. After optimization, a sequence of insulin doses, \textcolor{black}{$u_0,\,u_1,\,\hdots,\,u_{N_c-1}$}, is obtained. From this sequence, only the first input is delivered to the patient $u(t)=u_0$ and then the whole process is repeated after 5~minutes. The optimization can be understood as a function $u(t)=MPC(x(t))$.

In the following, the state-space model, the MPC optimization and the training data format are described in more detail.

\subsubsection{\textcolor{black}{Glucose-Insulin Prediction Model \cite{garcia2021advanced}}}
The following set of equations represents the prediction model:
\begin{equation}\label{eq:Model}
\begin{aligned}
    \dot{G}(t) &= -S_g[G(t)-G_{b}]-S_i\chi(t)G(t)+d(t),\\
    \dot{\chi}(t) &= -p_2\chi(t)+p_2\left[I_p(t)-I_{b}\right],\\
    \dot{d}(t) &= 0,\\
    \dot{I}_{sc1}(t) &= -(k_{a1}+k_d)I_{sc1}(t)+u(t),\\
    \dot{I}_{sc2}(t) &= -k_{a2}I_{sc2}(t)+k_dI_{sc1}(t),\\
    \dot{I}_p(t) &= -k_{cl}I_p(t)+\left[\frac{k_{a1}I_{sc1}(t)+k_{a2}I_{sc2}(t)}{V_IBW}\right],\\
\end{aligned}
\end{equation}
where $G$ (mg/dL) is the blood glucose concentration, $\chi$ (mU/L) is the insulin concentration in the remote compartment, $d$ (mg/dL/min) is a disturbance that accounts for unmodeled phenomena such as meal intakes, $I_{sc1}$(mU) and $I_{sc2}$(mU) are the amount of non-monomeric and monomeric insulin in the subcutaneous space, $I_p$ (mU/L) is the plasma insulin concentration, and $u$ (mU/s) is the rate of delivered insulin.

The model is linearized at $(u_{op}, G_{op})=(u_b,120)$, with $u_{b}$ a subject specific basal insulin rate set by the treating physician; and then it is discretized with a period $T=5$ (min) in order express it in the standard MPC linear form:
\begin{equation}\label{eq:model_linear}
\begin{aligned}
    x_{k+1}=Ax_k+B_uu_k+B_dd_k,\quad y_k = [1,0,0,0,0]x_k,
\end{aligned}
\end{equation}
where $x_k=[G_k,\,X_k,\,{I}_{sc1,k},\,{I}_{sc2,k},\,I_k]$ is the state vector, $d_k$ is the unknown disturbance, and $u_k$ (mU/min), $y_k$ (mg/dL), are  the insulin deliveries and glucose measurements.

\subsubsection{\textcolor{black}{Optimization problem \cite{garcia2021advanced}}}
The MPC optimization is:
\begin{equation}\label{eq:MPC_implicit}
    \begin{aligned}
    &\min_{u_i,\eta_i}  \sum_{k=0}^{Nc-1} \left[Q(IOB)\cdot(y_k-r_k)^2 + \kappa\cdot\eta_k^2 + \lambda(y_0,\dot{y}_0)\cdot u_k^2\right], \\
    & \hspace{1cm}\text{subject to:}\\
    & \hspace{1.75cm} \text{model \eqref{eq:model_linear}},\\
    & \hspace{1.75cm} -u_{b}\leq u_k\leq 1000-u_b,\\
    & \hspace{1.75cm} -50\leq u_{k}-u_{k-1}\leq 50,\\
    & \hspace{1.75cm} -50-y_k\leq \eta_k,\; \eta_k\geq 0,\\
    & \textcolor{black}{\hspace{1.75cm} r_k=\left\lbrace\begin{matrix} y_0\cdot e^{k/10} \hspace{0.5cm}\text{if $y_0\geq 0$}\\ 0 \hspace{1.6cm}\text{if $y_0<0$} \end{matrix}\right.}\\
    & \textcolor{black}{\hspace{1.75cm} d_k=\left\lbrace\begin{matrix} d_0 \hspace{0.55cm}\text{if $\dot{y}_0\geq0.05$}\\ \alpha_kd_0 \hspace{0.3cm}\text{if $\dot{y}_0<0.05$} \end{matrix}\right.}\\
    \end{aligned}
\end{equation}

The cost includes three main terms: \textbf{i)} $Q(IOB)\cdot(y_k-r_k)^2$, \textbf{ii)} $\kappa\cdot\eta_k^2$ and \textbf{iii)} $\lambda(y_0,\dot{y}_0)\cdot(u_k)^2$, respectively. The first one is responsible for penalizing deviations of the glucose measurements, $y_k=G_k$, from a reference signal that converges to the target $y_k=120$ (mg/dL). The second term penalizes hypoglycemia events through the soft constraint $-50-y_k\leq \eta_k$. The third term is a regularization penalizing deliveries of insulin.

The weighting parameter $Q(IOB)$ is set in terms of the Insulin On Board (IOB), a scalar that represents the remaining insulin in the body, see \cite{zisser2008bolus}. $Q(IOB)$ decreases as the IOB increases in order to allow aggressive controller action early in rejecting large disturbances (e.g. meals) while avoiding significant dosing later. Similarly, the parameter $\lambda(y_k,\dot{y}_k)$ becomes smaller when glucose rapidly increases. The goal is to make the MPC more aggressive when a rapid increase in glucose is observed and less aggressive when it is decreasing. \textcolor{black}{Finally, in order to to improve the glucose predictions that the MPC computes, the disturbance prediction is considered constant if the glucose rate of variation is positive, while it is smoothly driven to zero by $\alpha_i$ if the glucose is decreasing.}

\subsubsection{Training data format}
The MPC~\eqref{eq:MPC_implicit} can be understood as a deterministic input-output mapping of the form:
\begin{equation}
    u(t) = MPC(\tilde{x}(t)),
\end{equation}
where $u(t)$ is the first element of the optimal sequence of insulin deliveries obtained after solving the optimization, while $\tilde{x}(t)=[x(t),d(t),\dot{y}(t), IOB(t)]$ is an augmented-state vector containing all the parameters that are needed to solve the optimization. The variables $x(t)$ and $d(t)$ are computed with a Kalman filter, $\dot{y}(t)$ is obtained by numerical differentiation and $IOB(t)$ from the recent history of insulin delivery.

\begin{figure}[t]
    \centering
    \includegraphics[width=2.15in]{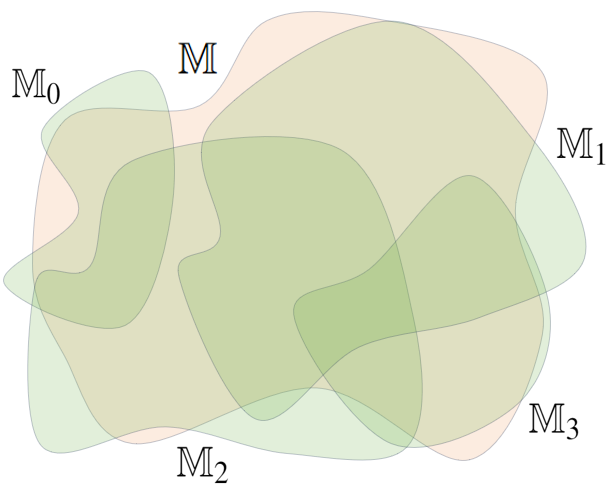}
    \caption{The MPC operational space (orange) vs. the one that is extracted in each randomized simulation (greens).}
    \label{fig:operation_domain}
\end{figure}

\subsection{Initial dataset generation}
In order to generate the initial dataset, $\mathcal{D}_o\approx \mathbb{M}$, the MPC algorithm is connected to the FDA-approved Type-1 Diabetes simulator developed at the University of Virginia~\cite{visentin2018uva, cobelli2023developing}. The goal is to simulate the resulting closed-loop under a vast set of meal and physical activity disturbances, providing us with data from $\mathbb{M}$. The simulator considered the following disturbances: meal's carbohydrates amount, meal's times, meal's absorption type (fast or slow), physical activity type, insulin type, insulin sensitivity, and rescue carbohydrates in case of hypoglycemia. Also, different modes of the control system were considered, such as fully/hybrid closed-loop. 

The idea is to run significant amounts of simulations in order to densely populate the MPC operational space, $\mathbb{M}$. To this end, a `day type' is first created by scheduling some disturbances at different time instants. For example, one day could be defined by 3 fast absorption meal disturbances of 50g of carbohydrates, appearing at 6am, 13pm and 19pm; two of them receiving manual pre-meal boluses. Then, a closed-loop simulation is performed, where the MPC handles the disturbances in closed-loop and delivers insulin to drive the blood glucose to its target of 120 mg/dL. That specific day provides data, $\{\tilde{x},\,u\}$, from a subdomain $\mathbb{M}_0\subseteq \mathbb{M}$ defined by the handled disturbances. 

The next day, a new set of randomized disturbances is created and simulated, providing data from a different subdomain $\mathbb{M}_1\subseteq \mathbb{M}$. The process is repeated indefinitely, ending up with a large set of data pairs that span across the operational domain defined by the simulation environment. This initial dataset $\mathcal{D}_0$ is an approximation of $\mathbb{M}$, however, it may be significantly unbalanced and therefore far from an OSD.

\begin{algorithm}[t]
\caption{Optimally sampled dataset}
\label{alg:OSD}
\begin{algorithmic}[1]
\State $\{J^*,\,S_x,\,S_u\} \gets \text{Set the optimal cost parameters.}$
\State $\mathcal{D} = \{X_0\}$ \Comment{Initialize final dataset with one random element from the initial set, $X_0\in\mathcal{D}_0$.}
\State $u_s=0$
\For {$X_k$ in $\mathcal{D}_0$}
    \State $X_{k}^{nn} \gets \text{get the $\mathcal{D}$-nearest neighbor of } X_k$.
    \If {$J(X_k, X_k^{nn}) > J^*$} \State $\mathcal{D} \gets X_k$ \Comment{Add $X_k$ to $\mathcal{D}$}
    \ElsIf {$|u_i-u_i^{nn}| > u_s$}
    \State $u_s \gets |u_i-u_i^{nn}|$ \Comment{Update $u_s$}
    \EndIf
\EndFor
\end{algorithmic}
\end{algorithm}
\subsection{An algorithm to build OSDs}
Algorithm~\ref{alg:OSD} was developed, which takes the initial set, $\mathcal{D}_0$, and returns a final smaller set, $\mathcal{D}$, satisfying the OSD conditions. The idea behind the algorithm is as follows: populate the final set $\mathcal{D}$ element by element, adding only elements from $\mathcal{D}_0$ that find a $\mathcal{D}$-neighbor with cost higher than~$J^*$. If the initial set $\mathcal{D}_0$ is sufficiently sparse, then this simple rule will guarantee the first and the second OSD conditions.

Concretely, the second condition is directly ensured by the `if' condition in line 6, which prevents the addition of new elements whose $\mathcal{D}$-nearest neighbor have cost lower than $J^*$. The first condition will be met as long as the initial dataset is sufficiently sparse. As the for loop progresses, the likelihood of satisfying the `if' condition in line 6 decreases. This happens because, as more elements are added to the final set $\mathcal{D}$, it becomes increasingly likely to find a sufficiently close neighbor that fails to satisfy the condition $J(X_k, X_k^{nn}) > J^*$ in the subsequent iterations. Thus, the probability of adding new elements to the dataset diminishes as the dataset grows. Eventually, a point is reached where nearly every new state $X_k$ finds a sufficiently close $\mathcal{D}$-nearest neighbor in the existing dataset. This behavior reflects the saturation property of the OSD definition, which can be now measured by how frequently the if condition in line 6 is triggered.

The computational bottleneck of Algorithm \ref{alg:OSD} appears in Line 5, where the $\mathcal{D}$-nearest neighbor of $X_k$ needs to be computed. From Definition~2, a brute force solution requires computing the cost between $X_k$ and all the other states in $\mathcal{D}$; resulting in an $O(N_d^2)$ complexity that is not acceptable for big datasets. We suggest employing algorithms based on Hierarchical Navigable Small World (HNSW) graphs to execute fast nearest-neighbor retrieval, which provide $O\big(log(N_d)\big)$ complexity and scale well to big volumes~\cite{malkov2018efficient}.

\section{Results}\label{sec:results}
We have generated 750 simulations of 60 days each for the 100 virtual subjects of the UVA Type-1 Diabetes simulator, summing up to 4.5 million days (12,328 years) of virtual data belonging to the MPC domain $\mathbb{M}$. 

\begin{figure*}[t]
    \centering
    \includegraphics[width=5.5in]{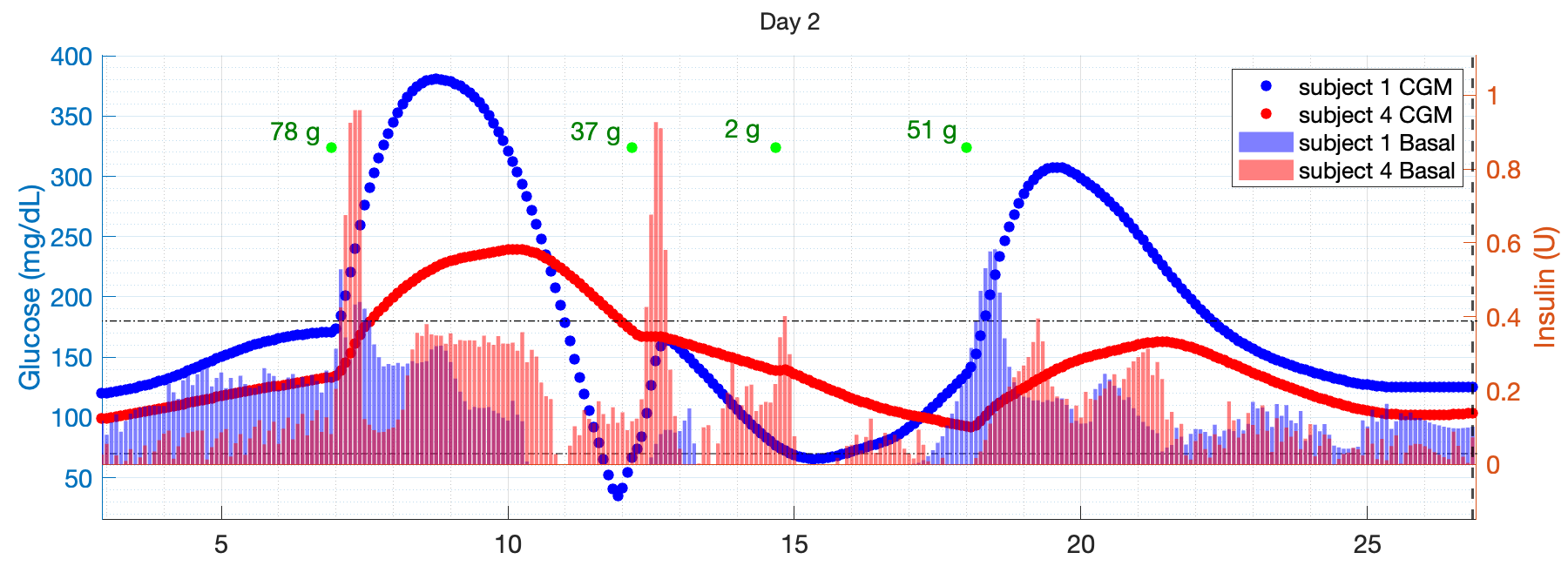}
    \caption{Illustration of one of the simulated days where four meal disturbances were given at different instants.}
    \label{fig:simulation_day}
\end{figure*}

Figure~\ref{fig:simulation_day} illustrates one of the simulated days, depicting the glucose-insulin evolution for two virtual subjects during one day. The UVA-MPC controller generated the basal insulin (red and blue bars). After the simulation, all the generated pairs $X_t=\{\tilde{x},u\}$ are stored in a database, which conforms the initial dataset $\mathcal{D}_0\approx \mathbb{M}$. As one data element is generated per each control cycle, the initial set contains about 1.3 billion elements.

\textcolor{black}{\subsection{Generating OSDs with different cost parameters}
We applied Algorithm~\ref{alg:OSD} to the large, unfiltered dataset $\mathcal{D}_0$, in order to create multiple OSDs. To this end, we performed a systematic grid search to test Algorithm~\ref{alg:OSD} under different combinations of cost-parameters. First, we normalized the state-distance metric by setting \(S_x\) to the inverse of the sample state covariance matrix (i.e. Mahalanobis distance). Then, we varied \(J^*\) over \(\{1,\,0.5,\,0.25,\,0.1\}\) and \(S_u\) over \(\{0,\,0.0001,\,0.0025,\,0.01\}\), generating multiple OSDs with different levels of granularity and numerical resolution.}

Recall that each combination of $J^*$ and $S_u$ creates a specific partition of the MPC operational space $\mathbb{M}$ in small volumes, refer to Figure~\ref{fig:D}. Hence, each state in the final OSDs creates an empty volume defined by Eq.~\eqref{eq:partition_volume}. By changing the values of $J^*$ and $S_u$, the volume boundaries are modified and therefore the final dataset granularity and numerical resolution change.

\subsubsection{\textcolor{black}{Effect of Cost Parameters on Volume Boundaries}}
Figure~\ref{fig:sampling_boundaries} illustrates the 2D projections of three partition volumes corresponding to different centroid states \(x_k\) in the final OSD. The background colormap shows the absolute difference between the MPC control action at the centroid and at neighboring states, which is zero at the centroid and increases with distance. \textcolor{black}{When \(S_u=0\), the volume boundary for each centroid is the ellipsoid defined by \(\,(x_k - x)^T S_x (x_k - x) = J^*.\) As \(S_u\) increases, this boundary is governed by
$(x_k - x)^T S_x (x_k - x) \;+\; (u_k - u)^T S_u (u_k - u)\;=\;J^*$,
thereby shrinking more aggressively in directions where the MPC output \(u\) is highly sensitive to changes in \(x\). This adaptive “volume compression” leads to denser sampling in critical areas and ultimately improves numerical resolution, as further discussed below.}

\begin{table}[t]
\centering
\begin{tabular}{|c|c|c|c|c|}
\hline
  & \textbf{\scriptsize $J^*$=1} & \textbf{\scriptsize $J^*$=0.50} & \textbf{\scriptsize $J^*$=0.25} & \textbf{\scriptsize $J^*$=0.1} \\ \hline

\textbf{\scriptsize \textbf{$S_u=0$}} & 
\begin{tabular}[c]{@{}c@{}}{\scriptsize $7.36\ (303.25)$} \\ {\scriptsize $N_d=11.71$}\end{tabular} & 
\begin{tabular}[c]{@{}c@{}}{\scriptsize $4.91\ (279.75)$} \\ {\scriptsize $N_d=26.24$}\end{tabular} & 
\begin{tabular}[c]{@{}c@{}}{\scriptsize $2.98\ (314.81)$} \\ {\scriptsize $N_d=53.03$}\end{tabular} & 
\begin{tabular}[c]{@{}c@{}}{\scriptsize $1.78\ (298.76)$} \\ {\scriptsize $N_d=114.36$}\end{tabular} \\ \hline

\textbf{\scriptsize \textbf{$S_u=0.0001$}} & 
\begin{tabular}[c]{@{}c@{}}{\scriptsize $4.48\ (57.46)$} \\ {\scriptsize $N_d=13.07$}\end{tabular} & 
\begin{tabular}[c]{@{}c@{}}{\scriptsize $2.95\ (40.57)$} \\ {\scriptsize $N_d=29.65$}\end{tabular} & 
\begin{tabular}[c]{@{}c@{}}{\scriptsize $1.79\ (28.59)$} \\ {\scriptsize $N_d=57.72$}\end{tabular} & 
\begin{tabular}[c]{@{}c@{}}{\scriptsize $1.06\ (19.59)$} \\ {\scriptsize $N_d=140.3$}\end{tabular} \\ \hline

\textbf{\scriptsize \textbf{$S_u=0.0025$}} & 
\begin{tabular}[c]{@{}c@{}}{\scriptsize $1.31\ (13.18)$} \\ {\scriptsize $N_d=18.13$}\end{tabular} & 
\begin{tabular}[c]{@{}c@{}}{\scriptsize $0.81\ (11.32)$} \\ {\scriptsize $N_d=39.35$}\end{tabular} & 
\begin{tabular}[c]{@{}c@{}}{\scriptsize $0.47\ (6.68)$} \\ {\scriptsize $N_d=74.94$}\end{tabular} & 
\begin{tabular}[c]{@{}c@{}}{\scriptsize -}\end{tabular} \\ \hline

\textbf{\scriptsize \textbf{$S_u=0.01$}} & 
\begin{tabular}[c]{@{}c@{}}{\scriptsize $0.68\ (6.45)$} \\ {\scriptsize $N_d=21.76$}\end{tabular} & 
\begin{tabular}[c]{@{}c@{}}{\scriptsize $0.61\ (6.99)$} \\ {\scriptsize $N_d=46.28$}\end{tabular} & 
\begin{tabular}[c]{@{}c@{}}{\scriptsize $0.28\ (3.26)$} \\ {\scriptsize $N_d=83.32$}\end{tabular} & 
\begin{tabular}[c]{@{}c@{}}{\scriptsize -}\end{tabular} \\ \hline
\end{tabular}
\vspace{0.25cm}
\caption{Observed numerical resolution and final dataset size for different combinations of $S_u$ and $J^*$. Top row: Mean and maximum control action difference, $u_s$, across all partition volumes. Bottom row: Final dataset size in millions.}
\label{table:1}
\end{table}

\subsubsection{\textcolor{black}{Effect of Cost Parameters on Numerical Resolution}}
\textcolor{black}{Table~\ref{table:1} illustrates how numerical resolution is affected by the cost parameters. Here, \emph{numerical resolution} refers to how well the OSD replicates the MPC control action when acting as a look-up table. The accuracy was measured as follows: for every state $x$ in the initial dataset $\mathcal{D}_0$, we identified its corresponding OSD partition-volume, and then we compared the true MPC control action (i.e. $u=MPC(x)$) with the one at the  volume centroid (i.e. $u_k=MPC(x_k)$, where $x_k$ is the $\mathcal{D}$-nearest neighbor of $x$). If this error is small for all $x$, then the OSD can be used as a look-up table for MPC implementation.}

\begin{figure*}[t]
    \noindent
    \includegraphics[width=\textwidth]{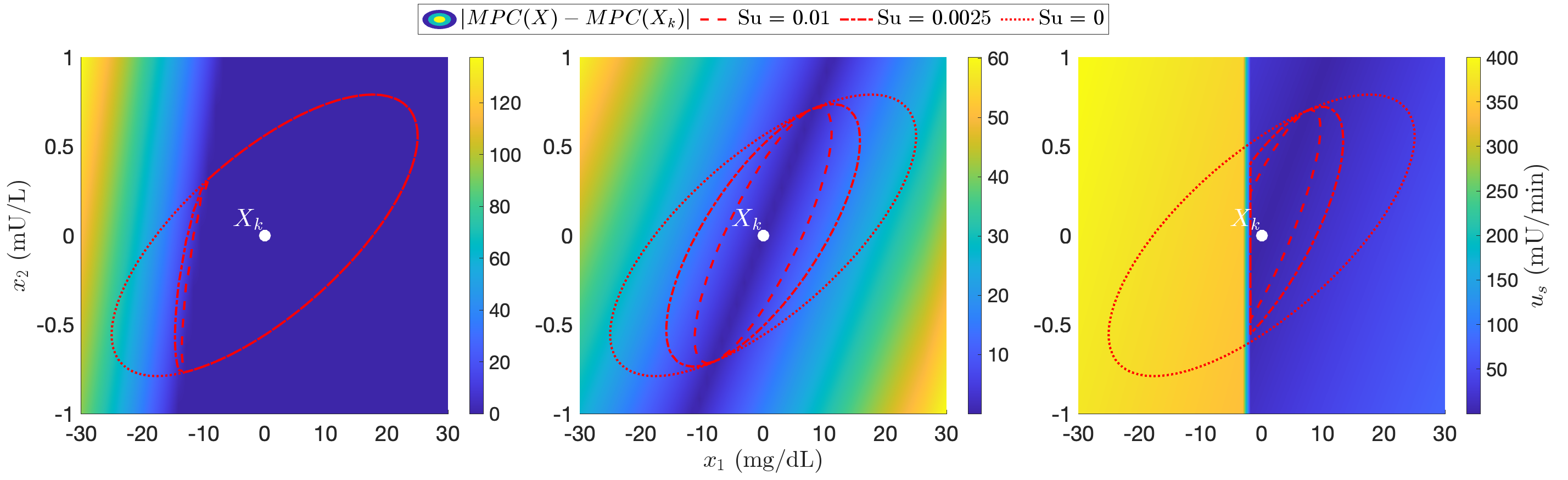}
    \caption{Two-dimensional projection of the partition-volume boundary for three different states. \textbf{Left:} the MPC is not sensitive in the positive direction of $x_1$, where the control action error remains close to zero. Increasing $S_u$ cuts the left part of the ellipsoid due to the $S_u$ penalization. \textbf{Middle:} the MPC is sensitive in the positive and negative directions of $x_1$. Increasing $S_u$ shrinks and aligns the ellipsoid to the lowest error direction. \textbf{Right:} The state that produced the highest error, i.e. $\max\{u_s\}=303.25$. The ellipsoid is shrunk in a non-symmetric way, fitting the lowest error area and reducing the error to $\max\{u_s\}=6.45$.}
    \label{fig:sampling_boundaries}
\end{figure*}

On the one hand, as \(J^*\) decreases, the average numerical resolution improves at the expense of significantly increasing the final OSD size. This happens because the ellipsoids defining the partition volumes uniformly shrink as \(J^*\) is made smaller, which reduces the probability of finding significantly different control actions within each volume. On the other hand, increasing \(S_u\) drastically reduces both average and maximum control-action errors at only a moderate cost in dataset size. This happens because, as illustrated in Figure~\ref{fig:sampling_boundaries}, the volumes adaptively shrink only in the directions that minimize control action error, which ensures that the final OSD allocates finer sampling where the MPC solution is most variable.

\begin{figure}[t]
    \noindent
    \centering
    \includesvg[width=3.5in]{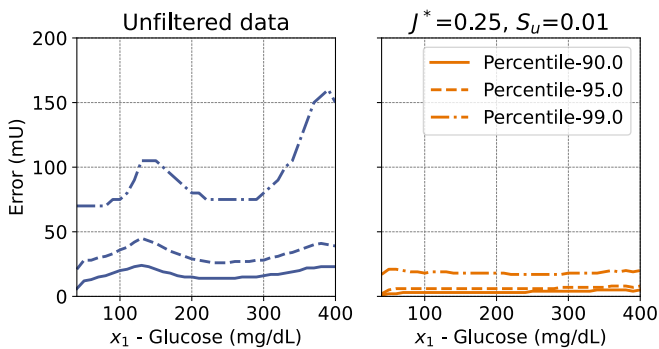}
    \caption{Comparison of the NN performance when trained on the raw data (left) vs. on an OSD (right). The Y-axis represents the error $|MPC(x)-NN(x)|$. The X-axis shows the distribution of the error across blood glucose levels.}
    \label{fig:nn_errors_fig_1}
\end{figure}

\subsection{Training MPC Neural Networks in OSDs}
A ResNET architecture (i.e. Figure~\ref{fig:ResNET}, middle) was trained on datasets generated with different \textcolor{black}{grid-search combinations of $J^*=\{0.25,\,0.5\}$ and $S_u=\{0,\,0.1\}$}. After training, a dataset generated with $J^*=0.1$ and $S_u=0$ was used for testing due to the benefits discussed latter in Section~\ref{sec:testing_datasets}.

\subsubsection{\textcolor{black}{Comparison between training in raw data versus training in OSDs}}
Figure~\ref{fig:nn_errors_fig_1} illustrates a projection of the control action error $u_s=MPC(x)-NN(x)$ in the blood glucose axis (i.e. the first state in~\eqref{eq:Model}). This axis is the most relevant in practice as the clinical metrics and the risks of long and short term complications are defined in terms of blood glucose. It can be seen that the NN-approximation error is significantly reduced and also flattened when trained on OSDs. As mentioned in Sec.~\ref{sec:training_sets}, the initial set has a skewed probability distribution that negatively affects learning. The OSD removes all the state repetitions without ultimately losing numerical resolution, translating into a more accurate and homogeneous learning of the MPC function \eqref{eq:MPCnn}.

\subsubsection{\textcolor{black}{Comparison between training in training in OSDs versus uniform samplers}}
\begin{figure}[t]
    \noindent
    \centering
    \includesvg[width=3in]{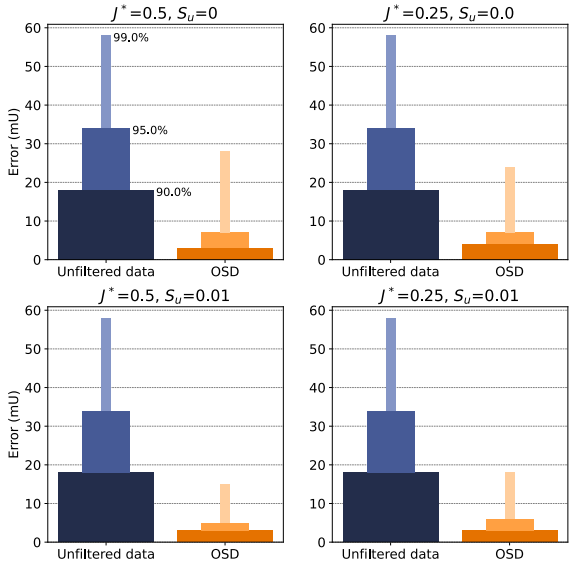}
    \caption{Comparison of the NN performance when trained on the raw data (blue) vs. on different OSDs (orange). The Y-axis represents the error $|MPC(x)-NN(x)|$ across all the testing data.}
    \label{fig:nn_errors_fig_2}
\end{figure}

Figure~\ref{fig:nn_errors_fig_2} compares the error when training on unfiltered data (blue) versus OSDs generated with different combinations of parameters (orange). \textcolor{black}{For comparative purposes, the datasets obtained with ${S_u=0}$ are equivalent to the ones that would be obtained by using the uniform sampling methods (i.e. hit-and-run algorithms) that have been proposed in previous studies \cite{winqvist2020training}. By definition, an OSD with $S_u=0$ is a uniform ellipsoidal sampling of $\mathbb{M}$. It can be seen how the penalization term $S_u>0$ helps to reduce the NN errors by 50\% in the case of $J^*=0.5$ and by 25\% for $J^*=0.25$.}

\subsection{Analysis of Computational Complexity}
The ResNET architecture has a total of 12 Residual Blocks (i.e. orange box in Fig.~\ref{fig:ResNET}), each of them containing two ReLu layers of size 16. The total size of the NN parameters was 41.02 kB, and it approximately needs 1 kB of available RAM memory to be executed (i.e. one layer loaded at a time). Computing the NN just involves to perform about 36 low-order matrix multiplications. In contrast, the MPC algorithm expressed in the standard quadratic form $y=\frac{1}{2}(x^T\cdot H\cdot x - c^T\cdot x)$, $A\cdot x\leq b$ has a matrix $H$ of size 48x48 and matrix $A$ of size 120x48. Solvers implementing the interior-point method need to solve the Karush-Kuhn-Tucker (KKT) conditions, which is an augmented system of linear equations with a square matrix of size 120+48=168. Solving this system with the Newton method may require a minimum of 110.25kB of RAM memory and execute several vector-matrix multiplications of order 168 in a process with multiple iterations. \textcolor{black}{The computational time needed by MPC was measured to be 180 times higher.}

\section{Discussion}\label{sec:discussion}
The results introduced in this paper can be used to create high-resolution training datasets for NN-MPC learning by using a simple and scalable cost-based mechanism. In this section, some elements of the method are further discussed.

\subsection{Initial dataset generation}
As mentioned in Sec.~\ref{sec:SODD}, the MPC operates on a subspace, $\mathbb{M}\subseteq\mathbb{R}^n$. In this paper, we used realistic simulators to extract data pairs $\{x(t),\,u(t)\}$ from the MPC operational space and create an initial dataset $\mathcal{D}_0\approx\mathbb{M}$. If this initial dataset is sufficiently sparse, then it will densely populate the space $\mathbb{M}$ and the Algorithm~\ref{alg:OSD} can then be used to filter it out.

However, it is also possible to employ randomized samplers instead of simulators (such as the hit-and-run algorithms \cite{Zabinsky2013}) to generate the initial set. These algorithms can create an uniform distribution of synthetic states across some predefined input region, which can be also filtered out with Algorithm~\ref{alg:OSD}.

\subsection{The saturation point}
The OSDs introduced here exhibit a 'saturation point,' which, as mentioned in Section~\ref{sec:OSD_alg}, is characterized by a continuous reduction in the probability of satisfying the `if' condition in line 6 of Algorithm~\ref{alg:OSD}.

Reaching saturation means that after some iteration `k', it becomes increasingly difficult to find states that trigger the `if' condition. Therefore, at saturation, no more states will be added to the final dataset. Saturation indicates the completeness of our initial training dataset; if observed, it implies that new simulations are not providing novel information (at least up to the numerical resolution specified by $J^*$).

In our results, we measured rejection ratios of 76\%, 87\%, 94\% and 98\% during the last 50 simulations processed by Algorithm~\ref{alg:OSD}, corresponding to $J^*$ values of 0.1, 0.25, 0.5 and 1, respectively. This suggests that our initial set of 1.3 billion elements contains almost all the necessary MPC information, and there was no need to generate additional data.

\subsection{Tuning rules for $J^*$, $S_x$ and $S_u$}
The cost parameters provide two distinct directions to optimize the filtering of the initial dataset.

On the one hand, reducing $J^*$ makes all the partition volumes proportionally smaller, contributing to increase the overall density but maintaining its uniformity. On the other hand, increasing~$S_u$ augments sampling in sensitive areas and it significantly helps to improve the numerical accuracy.

A simple tuning rule for $J^*$ and $S_u$ would be as follows: adjust first $J^*$ and $S_x$ in order to define a maximum-size ellipsoid that is reasonable for the problem at hand. Then run the algorithm with $S_u=0$ and check the obtained resolution. If the resolution is acceptable, then the process has finished. If it is not acceptable, then start increasing $S_u$ until you achieve the desired level of accuracy.

\subsection{Generation of testing datasets}\label{sec:testing_datasets}
If the NN has been trained on an OSD with parameters $\{J^*=J_{train},\;S_{u}=S_{train}\}$, then it is convenient to test its performance in a different OSD with $J^*<J_{train}$ and $S_u>S_{train}$.

The reason is that, by reducing $J^*$ and increasing $S_u$, the partition volumes are systematically made smaller; producing a testing OSD that has a finer granularity than the training one. By testing the NN in a more granular dataset, it will be systematically tested in system states $x$ that fall inside the empty volumes of the training set (i.e. between training states). This testing set helps to measure the interpolation capacity of the NN across all the states in the training set. If the same accuracy is observed with this testing set, it is an indication that the NN is correctly interpolating between elements.

\subsection{Inserting adaptive parameters or adaptive constraints}
It should be noted that if the MPC algorithm contains adaptive parameters, $\eta_k$, they can be included in an augmented state 
$\tilde{x}_k=[x_k,\eta_k]$ in order to make the NN also adaptive without the need for retraining.

Even with adaptive parameters, the MPC can still be understood as a deterministic function $u=MPC(x_k,\eta_k)=MPC(\tilde{x}_k)$. By considering the adaptive parameters as part of the input state, they become inputs to the NN as well, modifying the produced control action. Additionally, they are actively considered in the OSD filtering process.

This approach applies to: (i) adaptive parameters in the dynamic model $x_{k+1}=f(x_k,u_k,\eta_k)$, (ii) adaptive parameters in the cost functions $L(x,u,\eta)$, $M(x,\eta)$, or (iii) adaptive parameters in the MPC constrains $g(x_k, u_k, \eta_k) \leq 0$ and $h(x_k, u_k, \eta_k) = 0$. In our specific case, described in Section~\ref{sec:UVAMPC_description}, there were adaptive parameters in both the cost function and constrains.

\subsection{Limitations and application to other domains}
Although the reported results have been focused on insulin regulation for T1D, the methodology is general and therefore it may be applicable to other domains.

Any optimization problem that becomes computationally intensive with classical programming methods may be suitable for this technique. Normally, optimization algorithms need to run iterative loops with intense algebra in order to find a solution. NNs effectively eliminate the iterative process and therefore they can be used to retrieve the optimal solutions in a single forward pass without iterations.

In theory, the unique requirements for this technique to be applied to other domains are: i) the optimization problem can be expressed as $u=f(x)$, where $x$ represents the complete set of variables needed to solve the optimization, and $u$ is the optimization result; ii) there is capacity to artificially generate realistic target inputs $x_1,...,x_{N_d}$ and compute their associated outputs $u_1,...,u_{N_d}$; and iii) there is capacity to execute Algorithm~\ref{alg:OSD} without running into memory problems.

If those requirements are satisfied, then an initial dataset can be generated and filtered by Algorithm \ref{alg:OSD} to finally train NNs. Some problems that may appear in practice are:
\begin{itemize}
    \item \textbf{Lack of saturation:} Saturation is not observed after executing Algorithm~\ref{alg:OSD} and after maximizing the initial dataset size. In this case, it would be needed to reduce the numerical resolution until saturation is seen, and then check if the minimum resolution is sufficient. If it is not, then the operational input space may be excessively big to be properly sampled with a computer.
    \item \textbf{NN produces low accuracy}: The complexity of the optimization may result in an NN unable to learn the $u=f(x)$ relationship. In this case, increasing the number of NN parameters and making them deeper will help. Overall, it has been observed in prior work that ResNETs are more suitable for learning complex optimizations \cite{castillo2023deep}.
\end{itemize}

\section{Conclusion}
This work has introduced a cost-based methodology for building Optimally-Sampled Datasets (OSDs) that systematically capture the essential information required to train Neural Network surrogates of Model Predictive Controllers (MPCs). By adaptively refining the partitioning of the state space—focusing on regions where the controller is most sensitive—this method greatly enhances numerical resolution and ensures balanced coverage of important states.

Experiments conducted with the University of Virginia MPC algorithm for automated insulin delivery have shown significant improvements in the final NNs accuracy when they are trained on OSDs. Two NNs trained with these datasets received FDA Investigational Device Exemption and they were successfully tested in clinical trials with humans, representing the first time ever that an NN successfully delivered insulin to humans~\cite{kovatchev2024neural, NAP}. Future works could be directed in further exploring the benefits of this method for diabetes control as well as its applicability to other domains where efficient embedded optimization algorithms may be beneficial.

\section*{Conflict of interest}
Alberto Castillo, Elliot Pryor and Anas El Fathi hold Intellectual Property in the field of diabetes technology.

Boris Kovatchev has patent royalties handled by UVA LVG from: Dexcom, Lifescan, Novo Nordisk. Grants handled by UVA from Dexcom and Novo Nordisk. Study support from Tandem.

Marc Breton reports consulting and honoraria from BoydSense, Dexcom, Elivie, Roche, Portal Insulin, Tandem,  and Vertex;  patent royalties handled by UVA LVG from: Dexcom, Lifescan, Novo Nordisk. Grants handled by UVA from Dexcom, Novo Nordisk, and Tandem. In-kind research support from Dexcom and Tandem.

\bibliographystyle{IEEEtran}
\bibliography{biblio}

\end{document}